\documentclass[onecolumn,showpacs]{revtex4}
\usepackage{amsmath}
\usepackage{dcolumn}
\usepackage{graphicx}

\begin{document}

\title{A linearized spin-wave theory for thermodynamics of quantum Heisenberg antiferromagnet on a square lattice}
\author{M. M. Liang}
\author{Y. H. Su}
\email{suyh@ytu.edu.cn}

\affiliation{Department of Physics, Yantai University, Yantai
264005, China}
\date{\today }

\begin{abstract}

  The thermodynamics of the quantum Heisenberg antiferromagnet on a
  square lattice is revisited through a linearized spin-wave
  theory which is well defined at any finite temperature. We
  re-examine in details the temperature dependence of the free energy,
  the internal energy, the entropy and the specific heat. Most
  conclusions of the thermodynamics in previous studies can be
  reproduced in our linearized spin-wave theory. Specially, our
  calculation at low temperature $T<J$ agrees quantitatively with the
  numerical Quantum Monte Carlo simulation and high temperature series
  expansions.

\end{abstract}


\maketitle


The two-dimensional (2D) quantum magnets is one of the main
subjects in the modern condensed matter physics, as they are
closely related to the unconventional superconductivity in the
cuprates\cite{Manousakis} and the novel quantum spin-liquid states
in frustrated lattices.\cite{Nakatsuji} The previous studies are
mainly concentrated in the low-energy physics. However, some
interesting physics, such as the crossover behavior\cite{Kim1998}
in 2D quantum Heisenberg antiferromagnet (QHAFM) and the pseudogap
physics in the cuprates,\cite{Timusk} occurs at high energy scales
and are still in controversial. One reason for these controversies
is the lack of a reliable theoretical formalism well-defined in
full temperature region.

Recently we proposed a perturbation spin-wave theory for the 2D
QHAFM on a square lattice.\cite{Su} Our theory is based upon
Takahashi's modified spin-wave theory, where in order to study the
finite temperature disordered state, he introduced a constraint of
the loss of the local moment.\cite{Takahashi} It is found that the
2D quantum antiferromagnet is well described at low temperatures
by this modified spin-wave theory, yielding the results consistent
to the other theories\cite{ChakravartyPRL,AAPRL,AAPRB} and
numerical simulations.\cite{Takahashi,Makivic,Makivicb} In the
modified spin-wave theory, a mean-field ansatz is introduced to
decouple the spin-wave interactions. Although it gives a good
description of the low-energy physics, it results in the failure
of the theory at high temperatures $T\simeq J$, where an
artificial phase transition occurs which obviously violates the
Mermin-Wagner theorem.\cite{Mermin}

In our perturbation spin-wave theory, the loss of the local moment
is also included to fulfill the Mermin-Wagner theorem.  The
corrections from the spin-wave interactions are studied via a
many-body perturbation method, which removes the mean-field
divergence in the modified spin-wave theory. The temperature
dependence of the uniform static susceptibility from our
perturbation spin-wave theory agrees well with the quantum Monte
Carlo simulations and high temperature series expansions.\cite{Su}
It shows that our theory is well-defined at any finite temperature
and is a good formalism for the 2D QHAFM. Moreover it is found
that even the linearized spin-wave theory with the spin-wave
interactions neglected can reproduce the main features of the
uniform static susceptibility. Thus it encourages us to use the
linearized spin-wave theory, on its reliability and simplicity, to
study the thermodynamical properties of 2D QHAFM.

In this paper, we present our calculations of the thermodynamical
variables of a QHAFM on a square lattice, including the free
energy, the internal energy, the entropy and the specific heat.
Comparing with the numerical simulations, it shows that the
linearized spin-wave theory can also reproduce nearly all of the
features in thermodynamics. At low temperature $T<J$, our
calculations of the internal energy and the specific heat agree
quantitatively with the quantum Monte Carlo simulations and high
temperature series expansions.


 The Hamiltonian of a QHAFM on a square lattice is given by
\begin{equation}
H_{s} = J \sum_{\langle ij \rangle} \mathbf{S}_i \cdot
\mathbf{S}_j , \label{eqn2.1}
\end{equation}
where $J>0$ and $\langle ij\rangle$ denotes nearest-neighbor
sites. Since the ground state is a N\'{e}el ordered state, we
separate the lattice (with $2N$ sites) into two sublattice, where
sublattice A (B) is defined for the up (down) spins in the
N\'{e}el order state.

Following Takahashi, we introduce a Lagrange constraint into the
Hamiltonian,\cite{Takahashi}
\begin{equation}
H_{\lambda} = - \sum_{l\in A, B} \mu_l S_{l}^{z} . \label{eqn2.2}
\end{equation}
The lagrange multipliers are assumed as $\mu_l = \mu $ if $l \in A
$ and $-\mu$ if $l\in B$, and $\mu$ is re-defined as $\mu \equiv
JzS(\lambda -1)$ for convenience in our following discussion. This
Lagrange Hamiltonian fulfills the Mermin-Wagner theorem for the
loss of the local magnetization. It leads to a finite gap in the
spin-wave energy spectrum and thus the spin-wave excitations
behave as bosons with finite mass at finite temperatures.

In our spin-wave theory, the spin operators are expressed by
bosonic operators based upon the Dyson-Maleev
representation,\cite{Takahashi}
\begin{equation}
S_{i}^{-}=a_{i}^{\dag },S_{i}^{+}=(2S-a_{i}^{\dag
}a_{i})a_{i},S_{i}^{z}=S-a_{i}^{\dag }a_{i}, \label{eqn2.3-a}
\end{equation}%
for the spins in sublattice A and
\begin{equation}
S_{j}^{-}=b_{j},S_{j}^{+}=b_{j}^{\dag }(2S-b_{j}^{\dag
}b_{j}),S_{j}^{z}=-S+b_{j}^{\dag }b_{j},\label{eqn2.3-b}
\end{equation}%
in sublattice B. With this spin representation, the Hamiltonian
$H=H_{s} + H_{\lambda}$ can be expressed as summation of quadratic
and quantic terms in these bosonic operators. In our following
discussion, we will neglect the quantic terms which represent the
spin-wave interactions and investigate the thermodynamics of QHAFM
within a linearized spin-wave theory. The quadratic linearized
spin-wave terms in the Hamiltonian can be easily diagonalized
through the Fourier and Bogoliubov transformations. The free
energy per site is readily obtained as\cite{Su}
\begin{eqnarray} F
&=&\frac{T}{N}\sum_{\mathbf{k}}\ln \left( 2\sinh \left(
\frac{\varepsilon
_{\mathbf{k}}}{2T}\right) \right) -\frac{1}{2}Jz\lambda S(1+2S)  \notag \\
&&+\frac{1}{2} JzS^{2} ,  \label{eqn2.5}
\end{eqnarray}%
where the spin-wave spectrum $\varepsilon
_{\mathbf{k}}=JzS\sqrt{\lambda ^{2}-\gamma _{\mathbf{k} }^{2}}$
and $\gamma _{\mathbf{k}}=\cos \frac{1}{2}k_{x}\cos
\frac{1}{2}k_{y}$. $z=4$ is the coordinate number.

The parameter $\lambda$ is unknown and determined by the free
energy. Minimizing the free energy with respect to $\lambda $
leads to a self-consistent equation for it,\cite{Su}
\begin{equation}
S=-\frac{1}{2}+\frac{1}{2N}\sum_{\mathbf{k}} \coth \left( \frac{\varepsilon _{\mathbf{k}}}{%
2T}\right) \frac{\lambda }{\sqrt{\lambda ^{2}-\gamma
_{\mathbf{k}}^{2}}} .  \label{eqn2.6}
\end{equation}

\begin{figure}[tbp]
\includegraphics [angle=0,width=0.5\columnwidth,clip]{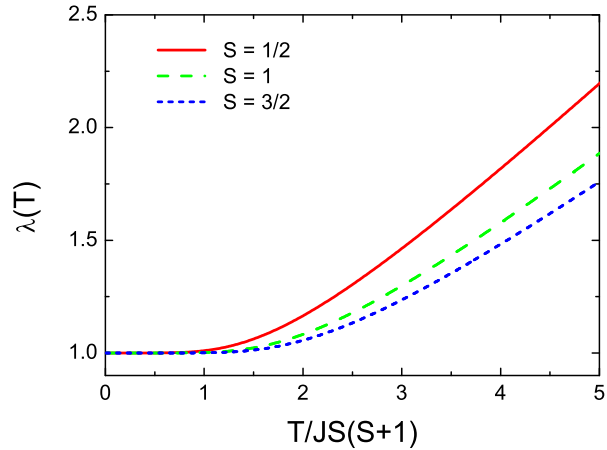}
\caption{The temperature dependence of the Lagrange multiplier
$\protect\lambda $ in our linearized spin-wave theory of QHAFM on
a square lattice with spins $S=1/2,1$ and $3/2$. } \label{fig2.1}
\end{figure}

The numerical calculation of the self-consistent equation
Eq.(\ref{eqn2.6}) has been present in details in our previous
paper.\cite{Su} Here we only cite the main results. The
temperature dependence of the Lagrange multiplier $\lambda$ is
shown in Fig. \ref{fig2.1}. At zero temperature the spin waves
condensate to the momentum $\left(0, 0\right)$ which results in a
long-range magnetic order.\cite{Takahashi,Hirsch} In this case,
$\lambda = 1 + O(1/N)$ and the spin-wave excitation is gapless. At
low but finite temperatures, the long-range order is destroyed by
the quantum fluctuations.\cite{Mermin} Then $\lambda$ has an
exponential behavior and decays to its zero-temperature value $1$
as $\lambda =1+\frac{1}{2}\left( \frac{T}{JzS}\right) ^{2}\exp
\left( -\frac{\pi
    JzSm_{0}}{T}\right)$.
Here $m_0 = S - 0.19660$ is the spontaneous ordered moment at zero
temperature. The spin waves thus behaves as bosons with  a finite
gap $\Delta = \sqrt{\lambda^2 -1}$. It is this finite gap that
leads to the vanishing of the long-range magnetic order at any
finite temperature. At high temperatures, the self-consistent
equation gives a weak dispersion for the spin waves with $\lambda
=\frac{T}{JzS}\ln \left( 1+\frac{1}{S}\right)$.

\begin{widetext}
\begin{center}
\begin{figure}[btp]
\includegraphics [angle=0,width=0.85\columnwidth,clip]{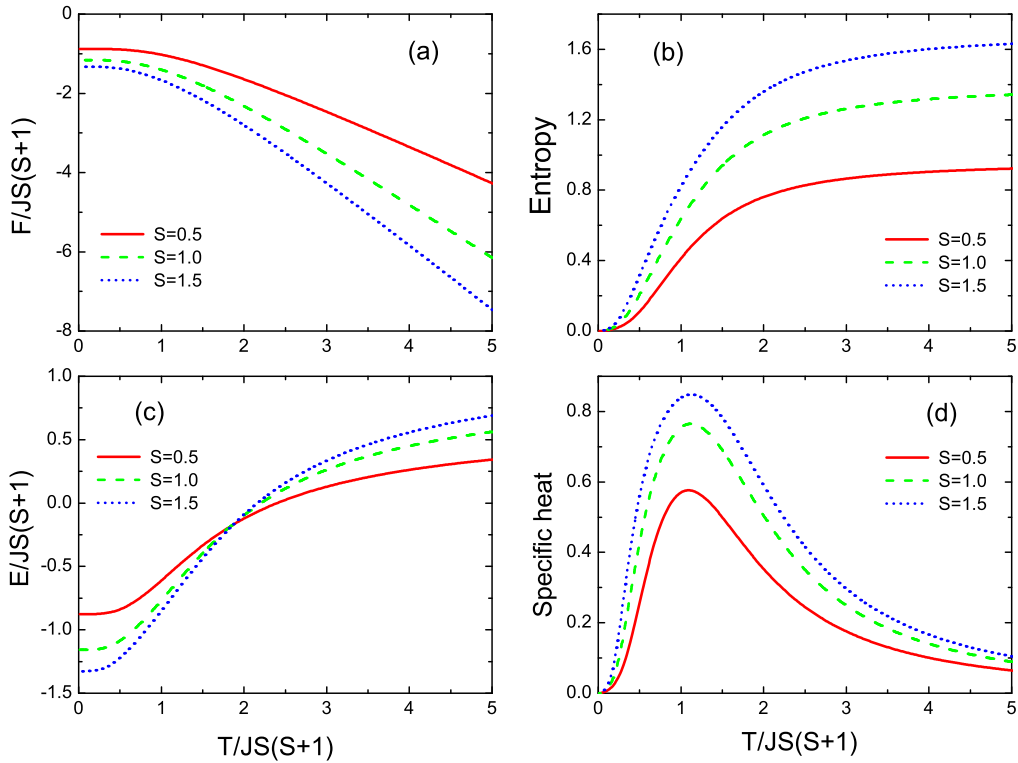}
\caption{Thermodynamics of the QHAFM on a square lattice with
different spin magnitudes, $S=1/2,1$ and $3/2$. (a) is the free
energy per site, (b) the entropy, (c) the internal energy and (d)
the specific heat.} \label{fig2.2}
\end{figure}
\end{center}
\end{widetext}

After the temperature variation of $\lambda$ is obtained, we have
calculated the uniform static magnetic susceptibility in our
previous paper.\cite{Su} Now let us turn our attention into the
thermodynamical properties. Fig. \ref{fig2.2} shows the
temperature dependence of the free energy, the entropy, the
internal energy and the specific heat of the QHAFM on a square
lattice. The free energy per site is obtained from
Eq.(\ref{eqn2.5}). In low temperature region,
\begin{equation*}
 F\left(T\right) = -\frac{1}{2}JzS\left(S+0.157948\right)+cT^3 .
\end{equation*}
where $c$ is a $S$-dependence constant. At high temperatures, it
decreases linearly with temperature,
\begin{equation*}
F = -T \ln \left[ \left(1+S\right)\left(1+S^{-1}\right)^{S}
\right] + \frac{1}{2} JzS^{2}.
\end{equation*}
This high-temperature behavior has been obtained by
Takahashi,\cite{Takahashi} where the last constant term does not
exist. Based on the good agreement of our internal energy and
specific heat with the numerical simulations as discussed below,
we argue that the finite constant term in our theory comes from
the linearized approximation.

The entropy shown in Fig. \ref{fig2.2} (b) is is calculated from
\begin{equation}
\mathcal{S}_n = \frac{1}{N} \sum_{\mathbf{k}} \left[ \left(
n_{\mathbf{k}}+1\right)\ln \left( n_{\mathbf{k}}+1\right) -
n_{\mathbf{k}}\ln n_{\mathbf{k}} \right] , \label{eqn2.9}
\end{equation}
where $n_{\mathbf{k}} =
\frac{1}{e^{\varepsilon_{\mathbf{k}}}/T-1}$. At zero temperature,
the entropy is zero, which is consistent to the second law of the
thermodynamics. It shows a square law $\sim T^2$ as temperature
increases. When temperature is high enough, it approaches a
constant value
\begin{equation*}
\mathcal{S}_n =\ln \left[
\left(1+S\right)\left(1+S^{-1}\right)^{S}\right] .
\end{equation*}
This value is different from $\ln\left(2S+1\right)$. It is also
due to the neglect of the spin-wave interactions in our theory.

\begin{figure}[tbp]
\includegraphics [angle=0,width=0.85\columnwidth,clip]{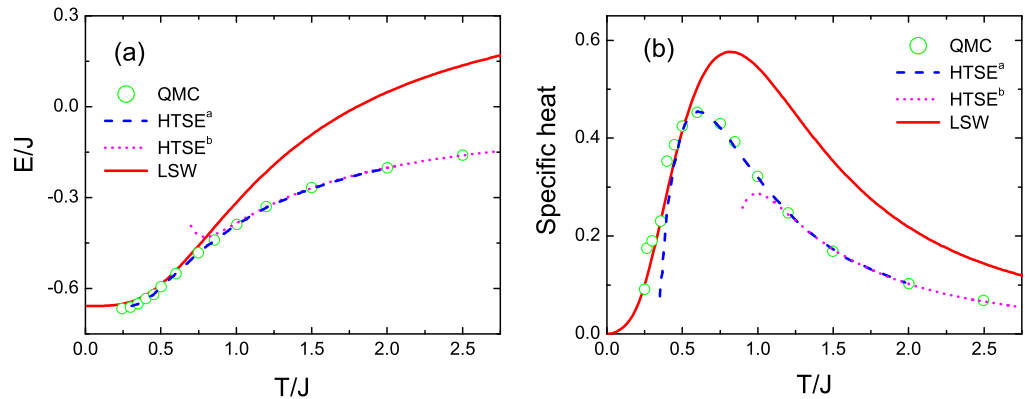}
\caption{Comparisons of the internal energy and the specific heat
from our linearized spin-wave theory (LSW) to the quantum Monte
Carlo (QMC) and high temperature series expansions (HTSE). Here
QMC data comes from Ref. [\onlinecite{Makivic}] and HTSE$^{a}$
from a method with Pad\'{e} approximation in Ref.
[\onlinecite{Wang1991}] and HTSE$^{b}$ from the formula Eq.(1) in
the same reference. } \label{fig2.3}
\end{figure}

The internal energy is calculated from the free energy and entropy
as $E = F + T \mathcal{S}_n$. At zero temperature for spin-1/2,
our calculation gives $E=-0.657948$. It is quite close to the
value $-0.669494$ which has included the spin-wave interaction
corrections,\cite{Zheng1992} and $-0.669437$ in quantum Monte
Carlo simulation.\cite{Sandvik1997} It shows that the spin-wave
interactions have a very weak contribution to the internal energy
at low temperatures. When temperature increases, the free energy
increases with an approximate  $T^{3}$-law and finally approaches
a constant value at high enough temperature. This asymptotic
behavior has also be shown in the high temperature series
expansions.\cite{Wang1991} A slight difference is that there is
constant value $E = \frac{1}{2}JzS^{2}$ in our theory, but it is
zero in high temperature series method. This difference stems from
our linearized approximation. A detailed comparison of the
internal energy from our linearized spin-wave theory with the
quantum Monte Carlo simulation and high-temperature series study
is shown in Fig. \ref{fig2.3} (a). A good agreement at temperature
$T<JS(S+1)$ shows that the spin-wave interactions has ignorable
contribution to the internal energy at low temperatures.

The specific heat can be calculated from $C_v = T\frac{\partial
\mathcal{S}_n}{\partial T}$ which leads to the following
expression,
\begin{equation}
C_v = \frac{1}{4N T^{2}} \sum_{\mathbf{k}}
\frac{\varepsilon_{\mathbf{k}}^{2} - \left(JzS\right)^{2}\lambda
T\frac{\partial \lambda}{\partial
T}}{\sinh^{2}\left(\varepsilon_{\mathbf{k}}/2T\right)} .
\label{eqn2.10}
\end{equation}
Here $\frac{\partial \lambda}{\partial T}$ can be obtained by the
derivative of both sides of the self-consistent equation
Eq.(\ref{eqn2.6}). The temperature dependence of the specific heat
is shown in Fig. \ref{fig2.2} (d). It vanishes at zero temperature
and shows an approximate  $T^{2}$-law at low temperatures. It
exhibits a broad peak at intermediate temperature $T\simeq
JS(S+1)$ and then decreases as $b T^{-2}$ at high temperature. A
more detailed analyses of our numerical data shows that $b=0.9075$
for $S=\frac{1}{2}$. Both the low temperature $T^{2}$ behavior and
$T^{-2}$ decay at high temperatures are consistent to the high
temperature series expansions.\cite{Bernu} The comparison with the
numerical simulations is given in Fig.  \ref{fig2.3} (b). It shows
that our linearized spin-wave theory has grasped main features in
the specific heat: a low temperature $T^{2}$ behavior, a smooth
crossover broad peak at intermediate temperature and a high
temperature $T^{-2}$ decrease.  A detailed comparison shows that
at low temperature $T<JS(S+1)$, our linearized spin-wave theory
agree quantitatively with the numerical simulations, which
indicates the validity of the spin-wave formalism in studying the
thermodynamics of the QHAFM.

\section{Summary}

In summary, we have calculated the thermodynamical quantities of
the QHAFM model on a square lattice from our linearized spin-wave
theory. Our calculations include the free energy, the entropy, the
internal energy and the specific heat. Although the neglect of the
spin-wave interactions in our theory leads to some difference in
magnitude compared to the numerical simulations, it has captured
main features of the thermodynamics. The quantitative agreement of
the internal energy and specific heat with the numerical
simulations at low temperature $T<JS(S+1)$ indicates that the
spin-wave theory is an appropriate formalism to study the 2D QHAFM
model.

\begin{acknowledgments}
We gratefully acknowledge valuable discussions with Dr. Fei Ye.
This work is supported by NSFC-China.
\end{acknowledgments}



\end{document}